# Study of Anisotropy on Ferromagnetic Electrodes of a Magnetic Tunnel Junction-Based Molecular Spintronics Device (MTJMSD)

Bishnu R. Dahal,[1] Marzieh Savadkoohi,[1] Eva Mutunga,[1] Rodneycia Taylor,[2] Andrew Grizzle,[1] Christopher D'Angelo,[1] and Pawan Tyagi[1*]

[1]Center for Nanotechnology Research and Education, Mechanical Engineering, University of the District of Columbia, Washington DC-20008, USA
[2]Department of Science, technology and Mathematics, Lincoln University, Jefferson City, MO-65101, USA

Corresponding Author: ptyagi@udc.edu

*Abstract*— **Magnetic tunnel junction-based molecular spintronics devices (MTJMSDs) are designed by covalently connecting the paramagnetic molecules across two ferromagnets (FM) electrodes of a magnetic tunnel junction (MTJ). MTJMSD provides opportunities to connect FM electrodes of a vast range of anisotropy properties to a variety of molecules of length scale. Our prior studies showed that the paramagnetic molecules can produce strong antiferromagnetic coupling with FM electrodes. The device properties of MTJMSD depend upon various factors such as anisotropy, spin fluctuation, thermal energy, etc. In this paper, we report a theoretical Monte Carlo Simulation (MCS) study to explain the impact of anisotropy on the MTJMSD equilibrium properties. We studied the energy variation of the MTJMSD system with time as a function of FM electrode anisotropy. Experimentally designed FM electrodes of MTJMSD contain multi-layers of different ferromagnetic materials. These materials possess in-plane and out-of-plane magnetic anisotropy characteristics. To understand the competing effect of in-plane and out-of-plane anisotropy, we have computationally applied anisotropies on the left FM electrode. For the MCS study, the orientation of the device was kept along YZ plane. As a result, the applied anisotropy along the X-direction ($A_{Lx}$) and Y-direction ($A_{Ly}$) represent out-of-plane and in-plane anisotropy, respectively. We found that increasing anisotropy strength starts exhibiting diverse domain structures within an FM electrode. Increasing the magnitude of anisotropy was found to create stripe-shaped domains with opposite spins. These domains represent the different magnetic phases. However, the application of equal magnitude of in-plane and out-of-plane cancels the strip domain formation and lowers the magnetic moment of overall MTJMSD**.

## I. INTRODUCTION

Molecular spintronics devices (MSDs) can overcome the miniaturization and heating issues associated with the existing computer technology[5]. These MSDs use the spin properties of the electrons and can revolutionize futuristic computational power. MSDs are also able to form the antiferromagnetic materials resulting from the molecule-induced strong Heinsburg exchange coupling[3, 8] between ferromagnetic (FM) electrodes and magnetic molecules. The antiferromagnetic spintronic devices provide many advantages e.g. they are robust against the external magnetic fields, capable of producing large magnetotransport properties, enabling the controlled spin-orbit coupling, etc [1] [4]. Given the nanoscale size of the molecules (∼ 2 nm), it is difficult to maintain the molecular dimension robust and reproducible gap between the two ferromagnetic leads[7]. To avoid these difficulties, we provide the new approach of making the magnetic tunnel junction-based molecular spintronics device (MTJMSD). To producing MTJMSD, the molecular channels were bridged across the insulator of an MTJ testbed with exposed side edges of FM electrodes. Device properties and their applications are highly influenced by various physical properties of the device such as various anisotropies, thermal energy, coupling of ferromagnetic atoms with its neighboring atoms or the coupling of ferromagnetic atoms with magnetic molecules, etc [6]. This manuscript mainly focuses on the Monte Carlo Simulation (MCS) study of MTJMSD when the right FM electrode is isotropic but in-plane and out-of-plane anisotropis are applied on the left FM electrode. All other parameters that can impact the overall magnetic properties of MTJMSD were kept constant during the Monte Carlo Simulation. Experimentally, bistable states in MTJMSD are realized by the utilization of two multi-layered magnetic electrodes of different magnetic properties deposited by the sputtering system. These multi-layered materials can possess different magnetic anisotropies. For example, an FM electrode with multi-layers of permalloy (nickel-iron magnetic alloy) and cobalt can be formed. Permalloy possesses in-plane anisotropy but cobalt provides the out-of-plane anisotropy[2]. It is a daunting task to understand the overall device properties of MTJMSD when individually in-plane and out-of-plane anisotropies are acting on two FM electrodes or competing effect of in-plane and out-of-plane anisotropies are acting on the single FM electrode. Our Monte Carlo Simulation systematically applied the in-plane and out-of-plane anisotropies individually and together on the left FM electrode. For keeping the discussion generic, the exchange coupling parameters, magnetic anisotropy, and thermal energy are referred to as the unitless parameters throughout this computational study.

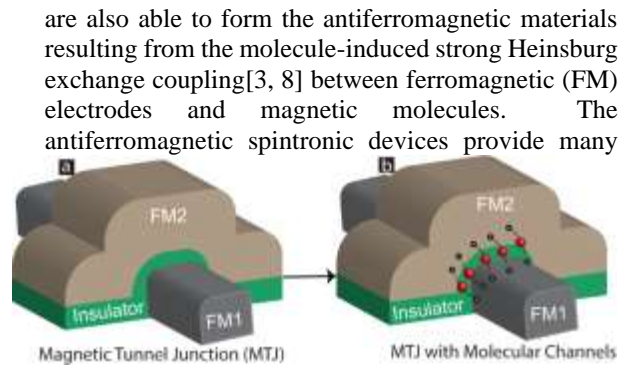

Figure 1. 3D sketch of MTJ (a) before and (b) after connecting molecular channels between two ferromagnetic electrodes.

## II. EXPERIMENT

Our MCS study investigates the overall magnetic properties of MTJMSDs when in-plane and out-of-plane anisotropies were applied on the same ferromagnetic (FM) electrode. To make our MCS study relevant to the experimentally observed MTJMSDs, we choose the multi-layers left FM electrode having in-plane and out-of-plane ansotropies while the right FM electrode is isotropic. Providing the experimental scenario, we have adopted the case when molecules produced antiferromagnetic coupling with one FM electrode and ferromagnetic coupling with another FM electrode. Magnetic tunnel junction (MTJ) is made of two cross junction of FM electrodes separated by a thin 2nm $nm$) insulator as shown in **Figure 1a**. Magnetic tunnel junction-based molecular spintronics devices (MTJMSDs) were created by connecting the molecular channels across the exposed FM electrodes as shown in **Figure 1b**. The schematic description of the dimension of MTJMSD including the spin orientation of molecules and FM atoms were desribed somewhere else[6, 7]. The Heisenberg coupling across the ferromagnetic atoms of left and right electrodes, represented by $J_L$ and $J_R$, always kept to their maximum values, i. e. $J_L = J_R = 1$ during the MCS. Similarly, $J_{mL}$ represents the Heisenberg coupling of molecules with the atoms of left FM electrodes while $J_{mR}$ represents the Heisenberg coupling of molecules with the atoms of right FM electrodes. To maintain the antiferromagnetic coupling of molecules with left and right FM electrodes, we always fixed the values of $J_{ML}$ = -1 and $J_{MR}$ =1. We studied the impact of unidirectional out-of-plane anisotropy along x-direction ($A_{Lx}$) and in-plane anisotropy along the y-direction ($A_{Ly}$) on the same left FM electrode. We varied all the possible combinations for $A_{Lx}$ and $A_{Ly}$ at thermal energy ($kT$) = 0.1. We varied the values of $A_{Ly}$ from no anisotropy ($A_{Ly}$ = 0) to its maximum value, i. e. $A_{Ly}$ = 1 for all possible values of $A_{Lx}$ (from $A_{Lx}$ = 0 to 1 at the step of 0.1). It is important to note that the value $kT = 0.1$ corresponding to the operational temperature (ranges from $50^0$C to $130^0$C of the device: close to room temperature) with the assumption of the Curie temperature of the FM varies from $500^0$C to $1300^0$C.

## III. RESULT AND DISCUSSIONS

**Figure 2** and **Figure 3** show the impact of anisotropies on the overall magnetic properties during the temporal evolution (Magnetic Moment Vs anisotropy) of the MTJMSD. Temporal evolutions were measured at $kT$ = 0.1. **Figure 2a** describes the variation of the magnetic moment of MTHMSD as a function of iteration counts when there is no anisotropy on the left FM electrode. Likewise, Figure 2b describes the the left-FM, and the right- FM electrodes can attain the maximum magnitude of the magnetic moment of 1250. Whereas MTJMSD maximum magnetic moment can settle around 2516 (1250 for each FM electrode and 16 for molecules). It is noteworthy to mention again that the right FM electrode is isotropic. As a result, the total magnetic moment of right FM electrode is always close to its maximum value of ~1200. When $A_{Ly}= 0$ and $A_{Lx} = 0$, the magnetic moment of the left FM electrode starts to increase linearly with the iteration counts. Magnetic moment of the left FM electrode saturates to its maximum value of ~1200 after 200 M iteration counts. In the absence of anisotropy, the antiferromagnetic coupling provided by the Heisenberg coupling of left and right FM electrode with the molecules are the dominating factor. Left and right FM electrodes were mentioned at antiferromagnetic coupling guided by the values of $J_{MR} = 1$ and $J_{ML} = -1$. The total magnetic moment of the MTJMSD is significantly small compared to that of the left and the right FM electrode, See **Figure 2a**, due to the opposite magnetic spins of the left and the right FM electrode. When $A_{Lx} = 0$ but $A_{Ly} = 0.5$, the in-plane anisotropy forced to reduce the magnetic moment of left FM electrode by overcoming the magnetic moment provided by $J_{ML}$, **Figure 2b.** This is possible because the in-plane anisotropy forced to alter the spin orientation of ferromagnetic atoms of left FM electrode provided by $J_{ML}$. It was also observed that the in-plane anisotropy caused the magnetic moment saturated right from the beginning of the iteration counts, **Figure 2b**. The impact of out-of-plane anisotropy and the competing effect of in-plane and out-of-plane anisotropy has described in **Figure 3a** and **Figure 3b**, respectively. **Figure 3a** demonstrates the situation having $A_{Ly} = 0$ and $A_{Lx} = 0.5$. The impact of out-of-plane anisotropy has a somewhat similar effect as provided with the equal magnitude of in-plane anisotropy. An interesting observation was noticed around 175M iteration counts. At this state, a sudden jump on the magnetic moment of the left FM electrode was observed. The magnetic moment saturates close to the value of the magnetic moment of ~425 immediately after the sudden jump, **Figure 3a**. With the application of in-plane and out-of-plane anisotropies on the same left FM electrode, we observed the competing effect. The competing impact of anisotropy helps to have a high value of magnetic moment by aligning all the magnetic spins of atoms of the left FM electrode. But the orientation of the magnetic spins of the left FM electrodes was opposite to that of the isotropic right FM electrode. As a result, the total magnetic moment of MTJMSD was observed to be significantly smaller than that of the left and the right FM electrodes. When the magnetic moment of left and right FM electrodes were closely equal but magnetic spins of left and right FM electrodes were opposite (~ 375 $M$ iteration counts), the total magnetic moment of MTJMSD was almost zero as demonstrated in **Figure 3b**. The overall magnetic moment of MTJMSD was similar when both FM electrodes were isotropic (**Figure 2a**) and when one FM electrode has an equal magnitude of in-plane and out-of-plane anisotropies (**Figure 4b**). The prior case was happened due to the antiferromagnetic Heisenberg coupling of the left and the right FM electrode with the paramagnetic molecule. While the latter case was due to the competing effect of in-plane and out-of-plane anisotropies on the left FM electrode.

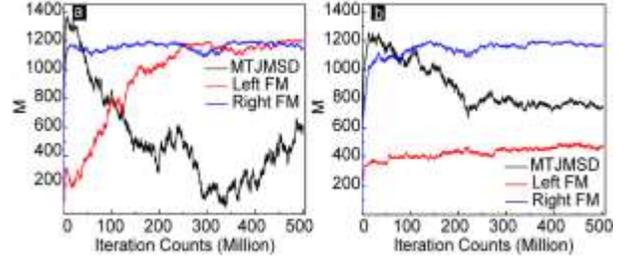

Figure. 2 Temporal evolution of the MTJMSDs device (Magnetic Moment Vs Iteration Counts) measured in terms of magnetic moment of MTJMSD, left FM and right FM electrodes for (a) variation of the magnetic moment as a function of iteration counts when in-plane anisotropy $A_{Ly} = 0.5$ but there is no out-of-plane anisotropy on the left FM electrodes. Based on the MTJMSD Heisenberg Model,

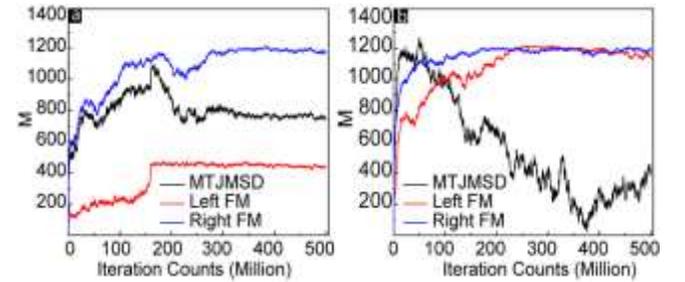

Figure. 3 Temporal evolution of the MTJMSDs device (Magnetic Moment Vs Iteration Counts) measured in terms of magnetic moment of MTJMSD, left FM and right FM electrodes for (a) $A_{Lx} = 0.5\ a$ and $A_{Ly} = 0$ and (b) $A_{Lx} = 1$ and $A_{Ly} = 1$.

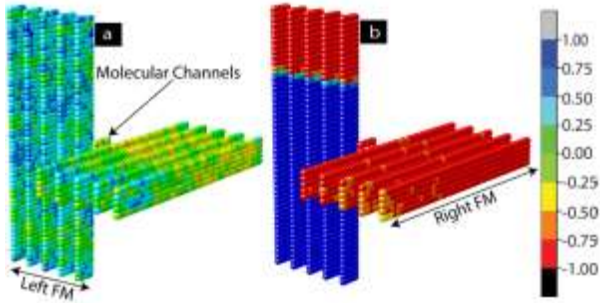

**Figure 4.** Simulated 3D lattice model of the MTJMSD measured at kT = 0.1 for (a) $A_{Lx}= 0$ and $A_{Ly} = 0$, (b) $A_{Lx}= 0$ $A_{Ly}= 0.5$

To understand the actual spin configurations of the left and right FM electrode, we analyzed the atomic scale equilibrium moment of MTJMSD's Heisenberg model, **Figure 4**. In 3D atomic schematic representation, the left FM electrodes are represented by vertical lattices while right FM electrodes are represented by horizontal lattices, and molecules are represented small square between left and right FM electrodes. The color scale bar presented in **Figure 4** represents the normalized magnetic moment along the X-spin direction. In the absence of anisotropy, magnetic spins were settled randomly either in the X, Y or Z -direction. But the application of anisotropy forced the magnetic spins to settle in the direction of the applied anisotropy. **Figure 4a** represents the 3D lattice model when both left and right FM electrodes are isotropic in nature. The Spin orientations are roughly random. However, if we carefully analyzed then one can observe that the net spin magnetic moment of left and right FM electrodes was opposite in nature. The net opposite magnetic spins were controlled by the antiferromagnetic Heisenberg coupling of the left and the right FM electrode with the paramagnetic molecules. This result totally agrees with the data presented in **Figure 2a**. Due to the application of in-plane anisotropy ($A_{Ly}$= 0.5) on the left FM electrode, multiple magnetic domains of opposite spins appeared on the left FM electrode. These domains of different magnetic spins actually represnt the different magnetic phases caused by the anisotropy on the single FM electrode. It is noteworthy to mention that despite having the isotropic right FM electrode, the spin orientations on the right FM electrodes were settled in a particular spin direction as shown in **Figure 4b**. Well settled spin states of the right FM electrode confirms that the net anisotropy present on the left FM electrode forced the spin waves to travel to the right FM electrode via the molecular channels. The study of the 3D lattice model of the competitive impact of in-plane and out-of-plane anistropies on the left, right, and both FM electrodes are the parts of the ongoing computational study of MTJMSD.

## IV. CONCLUSION

We have systematically studied the impact of anisotropy on the overall magnetic properties of magnetic tunnel junction-based molecular spintronics devices (MTJMSD). The MTJMDs were computationally simulated using Monte Carlo Simulation (MSD). We have applied the in-plane and out-of-plane anisotropies individually and simultaneously on the left ferromagnetic (FM) electrode while the right FM electrode is always isotropic. The temporal evolution of the MTJMSD shows that when there were no anisotropies, the total magnetic moment of the MTJMSDs was significantly small compared to that of the left and the right FM electrode. The application on in-plane or out-of-plane anisotropy forced to align the magnetic spins to a particular direction so that the magnetic moment of the MTJMSD would increase. The application of an equal amount of in-plane and out-of-plane anisotropy on the same left FM electrode cancels their mutual effect. The study of the 3D lattice model demonstrates the application of anisotropy on the left FM electrode starts to generate multiple magnetic phases of opposite spins on the left FM electrode. The competing effect of in-plane and out-of-plane anisotropies on the same electrode faded the multiple magnetic phases.

## ACKNOWLEDGMENT

This research is supported by National Science Foundation-CREST Award (Contract # HRD-1914751), Department of Energy/ National Nuclear Security Agency (DE-FOA-0003945).